\documentclass[sigconf, 9pt,]{acmart}
\usepackage{xcolor, soul}
\usepackage{amsthm}
\usepackage{caption}
\usepackage{subcaption}
\usepackage[inline]{enumitem}
\usepackage[export]{adjustbox}
\usepackage{xspace}
\usepackage{blindtext}
\usepackage{tabularray}

\usepackage{todonotes}
\usepackage{tikz}
\usepackage{hyperref}
\usepackage{multirow}
\usepackage{multicol}
\usepackage{listings}

\UseTblrLibrary{booktabs} 
\setcopyright{none}
\renewcommand\footnotetextcopyrightpermission[1]{} 
\AtBeginDocument{%
  }

\makeatletter
\DeclareRobustCommand{\onedot}{\futurelet\@let@token\@onedot}
\def\@onedot{\ifx\@let@token.\else.\null\fi\xspace}
 
\def\etc{\emph{etc}\onedot}

\makeatother

\newcommand{\origen}{\text{OriGen}\xspace}

\definecolor{bluea}{RGB}{232,237,250}
\definecolor{blueb}{RGB}{199,208,250}
\definecolor{bluec}{RGB}{164,182,250}
\NewTableCommand\h   {\SetCell{font=\bfseries}}
\NewTableCommand\hi  {\SetCell{font=\bfseries,bg=bluec}}
\NewTableCommand\hii {\SetCell{bg=blueb}}
\NewTableCommand\hiii{\SetCell{bg=bluea}}
\NewTableCommand\cii {\SetCell[c=2]{c}}
\NewTableCommand\ciii{\SetCell[c=3]{c}}
\NewTableCommand\rii {\SetCell[r=2]{c}}
\NewTableCommand\riii{\SetCell[r=3]{c}}
\NewTableCommand\riv {\SetCell[r=4]{c}}
\NewTableCommand\rv  {\SetCell[r=5]{c}}
\NewTableCommand\rvi {\SetCell[r=6]{c}}
\SetTblrInner{rowsep=0.5pt,colsep=2pt}
\begin{document}

\title{\origen: 
Enhancing RTL Code Generation with Code-to-Code Augmentation and Self-Reflection
}
\author{Fan Cui}

\affiliation{
  \institution{Peking University}
  \country{}
}
\email{pku_cf@stu.pku.edu.cn}

\author{Chenyang Yin}
\affiliation{%
  \institution{Peking University}
  \country{}
}
\email{ycy@stu.pku.edu.cn}

\author{Kexing Zhou}
\affiliation{%
  \institution{Peking University}
  \country{}
}
\email{zhoukexing@pku.edu.cn}

\author{Youwei Xiao}
\affiliation{%
  \institution{Peking University}
  \country{}
}
\email{shallwe@pku.edu.cn}

\author{Guangyu Sun}
\affiliation{%
  \institution{Peking University}
  \country{}
}
\email{gsun@pku.edu.cn}

\author{Qiang Xu}
\affiliation{%
  \institution{The Chinese University of Hong Kong}
  \country{}
}
\email{qxu@cse.cuhk.edu.hk}

\author{Qipeng Guo}
\affiliation{%
  \institution{Shanghai AI Laboratory}
  \country{}
}
\email{qpguo16@fudan.edu.cn}

\author{Demin Song}
\affiliation{%
  \institution{Shanghai AI Laboratory}
  \country{}
}
\email{songdemin@pjlab.org.cn}

\author{Dahua Lin}
\affiliation{%
  \institution{Shanghai AI Laboratory}
  \country{}
}
\email{lindahua@pjlab.org.cn}

\author{Xingcheng Zhang}
\affiliation{%
  \institution{Shanghai AI Laboratory}
  \country{}
}
\email{zhangxingcheng@pjlab.org.cn}

\author{Yun (Eric) Liang}
\authornote{Corresponding author}
\affiliation{%
  \institution{Peking University}
  \country{}
}
\email{ericlyun@pku.edu.cn}

\begin{abstract}
Recent studies have demonstrated the significant potential of Large Language Models (LLMs) in generating Register Transfer Level (RTL) code, with notable advancements showcased by commercial models such as GPT-4 and Claude3-Opus. 
However, these proprietary LLMs often raise concerns regarding privacy and security. 
While open-source LLMs offer solutions to these concerns, they typically underperform commercial models in RTL code generation tasks, primarily due to the scarcity of high-quality open-source RTL datasets.
To address this challenge, we introduce \origen, a fully open-source framework that incorporates self-reflection capabilities and a novel dataset augmentation methodology for generating high-quality, large-scale RTL code. 
Our approach employs a code-to-code augmentation technique to enhance the quality of open-source RTL code datasets. 
Furthermore, \origen can rectify syntactic errors through a self-reflection process that leverages compiler feedback.

Experimental results demonstrate that \origen significantly outperforms other open-source alternatives in RTL code generation. 
It surpasses the previous best-performing open-source LLM by 12.8\% and even exceeds GPT-4 Turbo in the pass@1 metric on the VerilogEval-Human benchmark.
Moreover, \origen exhibits superior capabilities in self-reflection and error correction, outperforming GPT-4 by 19.9\% on a benchmark designed to evaluate self-reflection capabilities. 

\textit{\origen is open source at GitHub(\href{https://github.com/pku-liang/OriGen}{https://github.com/pku-liang/OriGen})} 
\end{abstract}

\maketitle
\pagestyle{plain}

\section{Introduction}
Recent advancements in large language models (LLMs) have demonstrated exceptional capabilities in natural language comprehension and generation ~\cite{ammus}. 
Studies have shown that LLMs exhibit considerable proficiency in code generation tasks ~\cite{codesurvey}. 
Commercial LLMs, such as GPT-4~\cite{gpt4} and Claude3-Opus~\cite{claude3}, have showcased their ability to generate high-quality software code for common programming languages like C++ and Python, significantly enhancing coding productivity.
Moreover, LLMs have demonstrated impressive capabilities in hardware code generation, particularly for Register Transfer Level (RTL) code. 
The generation of RTL code from natural language instructions presents an innovative approach to enhance hardware development productivity. 
This method has the potential to revolutionize existing Hardware Description Language (HDL) coding workflows by alleviating the burdensome task of HDL coding for designers. 
Similar to High-Level Synthesis (HLS), LLMs can streamline the design process by enabling code generation from high-level specifications, thus simplifying complex hardware design tasks~\cite{canis2011legup, coussy2009introduction}.

While commercial LLMs have demonstrated proficiency in generating RTL code, they often raise concerns regarding privacy and security. 
These issues are particularly critical in the domain of hardware design, as RTL code frequently contains valuable intellectual property. 
Furthermore, the closed-source nature of these LLMs restricts researchers from conducting in-depth analyses and customizations, impeding further fine-tuning of the models in specific domains, such as hardware design. 
In contrast, open-source models provide enhanced privacy and security while facilitating further improvement and customization.
Open-source models, such as StarCoder2~\cite{stackv2}, DeepSeek-Coder~\cite{deepseek}, and CodeQwen-V1.5~\cite{qwen}, have demonstrated promising results in code generation for prevalent programming languages like Python and C/C++. 
However, their performance in RTL code generation still falls behind GPT-3.5. 
This performance gap underscores the pressing need to develop a specialized open-source LLM tailored for RTL code generation.

Nonetheless, there is a notable scarcity of high-quality, large-scale RTL code datasets compared to those available for other popular programming languages. 
This shortage poses a significant challenge in achieving satisfactory results for RTL generation tasks.
To address this issue, several studies have focused on collecting open-source code snippets ~\cite{benchmarking, verigen, betterv, deepframework}.
However, these open-source datasets may include low-quality code that can impair model performance~\cite{verilogeval}.
Alternative approaches have been developed to synthesize RTL code using LLMs ~\cite{verilogeval, dataisallyouneed, rtlcoder}.
VerilogEval~\cite{verilogeval} generates datasets by adding descriptions to open-source Verilog code but suffers from overall low quality. 
RTLCoder ~\cite{rtlcoder} selects hardware-related keywords as seeds to generate natural language instructions, which are then converted into corresponding RTL code using GPT-3.5. 
While this dataset meets quality standards, its scale is limited due to the restriction of generated data to selected keywords.

Self-reflection is another critical factor in RTL code generation. 
In this context, self-reflection refers to the model's ability to assess its generated code based on feedback from hardware compilers and simulators. 
This process involves identifying error causes and refining results accordingly, mirroring the hardware design process where RTL code undergoes rigorous evaluation to ensure correctness and adherence to design specifications.

Existing open-source LLMs primarily focus on code generation, often neglecting the interactive process of compilation and simulation integral to hardware design methodologies. 
To address these limitations, several studies have integrated self-reflection methodologies into their frameworks ~\cite{rtlfixer, autochip, chateda}. 
However, both RTLFixer~\cite{rtlfixer} and AutoChip~\cite{autochip} rely on commercial LLMs for self-reflection, raising potential privacy and security concerns as previously discussed.
Open-source LLMs generally exhibit weaker self-reflection capabilities compared to their commercial counterparts, primarily due to limited training data. 
Consequently, there is a pressing need to develop methodologies for constructing datasets specifically designed to enhance the self-reflection abilities of open-source LLMs in the context of RTL code generation.

In this paper, we introduce \origen, a fully open-source framework for RTL generation featuring self-reflection capabilities and a dataset augmentation methodology.
We propose a novel code-to-code augmentation technique to improve the quality of open-source RTL code datasets, which are typically large-scale but of limited quality due to the lack of rigorous review and validation processes.

Our approach involves extracting high-level code descriptions from open-source RTL code and refining the code based on these descriptions using the Claude3-Haiku model as a teacher. 
Commercial LLMs, trained on vast amounts of high-quality data, possess strong programming language understanding and generation capabilities. 
By leveraging these models as teachers, we can distill their knowledge to refine open-source RTL code, potentially achieving performance superior to the teacher model through rigorous review and validation.

TTo evaluate the LLM's self-reflection capability in RTL code, we construct a benchmark named VerilogFixEval, comprising 221 cases of failed compilation from VerilogEval~\cite{verilogeval}. 
\origen can correct syntactic errors by leveraging a self-reflection process based on compiler feedback. 
When generated code fails to pass compiler verification, the model initiates the self-reflection process, accepting erroneous code and compiler error messages to fix the RTL code, thereby enhancing its correctness and reliability.
The model's self-reflection ability is facilitated by a carefully constructed error-correction dataset, which includes natural language instructions, erroneous code, compiler error messages, and corrected code generated by Claude3-Haiku. 
This dataset serves as a valuable resource to bridge the gap in self-reflection capabilities between open-source and commercial closed-source LLMs. 
By training on this dataset, \origen can acquire self-reflection abilities comparable to those of commercial LLMs, enabling more reliable and accurate RTL code generation.

Our contributions are summarized as follows:
\begin{itemize}
    \item We introduce \origen, which significantly outperforms other alternatives designed for RTL code generation and achieves performance comparable to the latest version of GPT-4 Turbo. \textit{\origen is open source at \href{https://github.com/pku-liang/OriGen}{https://github.com/pku-liang/OriGen}}.
    \item  We propose a novel code-to-code augmentation methodology for generating high-quality, large-scale RTL code datasets, enhancing the model's training data.
    \item  We introduce a self-reflection mechanism that enables \origen to autonomously fix syntactic errors by leveraging compiler feedback, improving its code generation accuracy. 
    Furthermore, we construct a dataset to improve the model's self-reflection capability based on compiler error messages and develop a benchmark to evaluate this capability.
\end{itemize}

Experimental results demonstrate that \origen significantly outperforms other open-source alternatives in RTL code generation~\cite{betterv, rtlcoder, multi, verilogeval, chipnemo}. 
It surpasses the previous best-performing open-source LLM by 12.8\% and even exceeds GPT-4 Turbo in the pass@1 metric on the VerilogEval-Human benchmark.
Furthermore, \origen exhibits superior capabilities in self-reflection and error correction, outperforming GPT-4 by 19.9\% on the benchmark designed to evaluate self-reflection capabilities.


\section{Background and Related Work}
\subsection{LLMs for Verilog Generation}
Large Language Models (LLMs) have emerged as powerful tools for code generation across various programming languages. 
Trained on vast amounts of code and natural language data, these models can learn the statistical patterns and relationships within the training data, enabling them to generate code that adheres to the syntax and style of the target programming language.
Recently, researchers in the field of hardware design have shown a growing interest in leveraging LLMs for generating RTL code. 
RTL is a crucial abstraction level in hardware design, describing the flow of data and control signals between registers and combinational logic. 
By fine-tuning LLMs on RTL code datasets, models are capable of generating RTL code, potentially accelerating the hardware design process.

Among the pioneering endeavors, DAVE ~\cite{dave} represents an initial exploration into the generation of Verilog code from natural language instructions through the fine-tuning of the GPT-2 model. 
Subsequently, recognizing the notable scarcity of large-scale Verilog datasets relative to more common programming languages, Verigen ~\cite{verigen} collects a comprehensive Verilog dataset from GitHub's open-source repositories and various textbooks.
However, the performance of its optimally fine-tuned model based on the collected dataset remains inferior to that of GPT-3.5~\cite{rtllm}. 
This can be attributed to the varying quality of the collected open-source code, which was not filtered and processed.

Both VerilogEval ~\cite{verilogeval} and RTLCoder ~\cite{rtlcoder} acknowledge the importance of dataset quality and adopt data synthesis methods using commercial LLMs to improve the quality of their datasets.
VerilogEval proposes a synthesis approach that utilizes GPT-3.5 to add descriptions to each code snippet in the filtered VeriGen ~\cite{verigen} dataset.
This approach allows the LLM trained on the dataset to learn not only the syntax of Verilog but also to understand the correspondence between Verilog code and natural language. 
However, due to the average low quality of code in the dataset, VerilogEval still falls short of GPT-3.5.
RTLCoder~\cite{rtlcoder} adopts a synthesis approach to generate RTL code from natural language specifications. 
It selects one or two keywords from a prepared list of hundreds of digital design keywords and utilizes GPT-3.5 to generate RTL design instructions based on these keywords. 
Subsequently, GPT-3.5 is employed again to generate corresponding RTL code from these instructions. However, the reliance on a limited set of keywords poses challenges for further expansion of this dataset, as it may restrict the scale and complexity of the generated RTL code.
In contrast, the dataset synthesized through the code-to-code approach is not only of high quality, as demonstrated by the experimental results, but also extensive in scale, encompassing a diverse range of RTL code samples. 
This approach overcomes the limitations of above synthesis and enables the generation of a comprehensive and high-quality dataset that can effectively support the training of LLMs for RTL code generation.

Apart from fine-tuning open-source models, several studies ~\cite{chipchat, chipgpt, gpt4aigchip} investigate the direct application of commercial LLMs for generating RTL code. Chip-Chat ~\cite{chipchat} employs a conversational interface to design and verify an 8-bit accumulator-based microprocessor using GPT-4. ChipGPT ~\cite{chipgpt} introduces a four-stage, zero-code logic design framework that leverages GPT model.
\subsection{Reflection for Verilog Generation}
Existing open-source LLMs designed for RTL code generation have mainly focused on the generative aspects, neglecting the crucial stages of verification and reflection that are integral to hardware design methodologies. 
The effectiveness of self-reflection and receptivity to feedback in addressing real-world issues has been demonstrated by SWE-agent~\cite{sweagent}, which successfully resolved problems in GitHub's repositories.
This highlights the potential benefits of incorporating self-reflection mechanisms into LLMs for RTL code generation, as it could enable the models to learn from feedback and iteratively improve the generated code, aligning with the rigorous verification processes inherent to hardware design methodologies.

Recognizing the limitations of previous research on LLMs for RTL generation, AutoChip ~\cite{autochip} and RTLFixer ~\cite{rtlfixer} decide to introduce self-reflection into the their framework.
Specifically, RTLFixer employs GPT to generate and reflect on code. 
It includes approximately 80 erroneous code snippets that fail compilation, each paired with guidance. 
During the RTL code generation process, if the generated code fails to compile, RTLFixer ~\cite{rtlfixer} queries the database for the guidance most closely related to the error.
However, the generalization of this approach is limited due to the constraint of the guidance database.
In contrast, AutoChip ~\cite{autochip} directly provides the compiler error messages to the LLM rather than querying a database for guidance.

The error-correction dataset we construct is diverse and not limited to a specific set of errors, enabling the models to handle a broader range of issues encountered during RTL code generation. 
By leveraging this dataset, the models can improve their capabilities of self-reflection.




\subsection{Hardware Programming}
Hardware description languages such as Verilog or VHDL are widely adopted for hardware design. As an alternative, High-Level Synthesis (HLS) is a crucial methodology for hardware design~\cite{canis2011legup, coussy2009introduction}.
Similar to using LLMs to write RTL code, HLS enables designers to specify hardware functionality using high-level programming languages like C, C++, or SystemC.
These specifications are then automatically converted into hardware description languages such as VHDL or Verilog~\cite{xu2022hector, dai2019improving, josipovic2018dynamically, hsiao2019thread}. By automating this translation process, HLS simplifies the design of complex hardware systems, reducing the time and effort required to manage intricate low-level coding, allowing designers to optimize and manipulate designs more efficiently~\cite{ jia2021tensorlib, jia2022ems}.

\section{Methodology}
In this section, we provide comprehensive details of \origen, a powerful framework designed to enhance Verilog code generation.

\subsection{Overview} \label{subsec:overview}
\begin{figure*}[t]
  \centering
    \includegraphics[width=\textwidth]{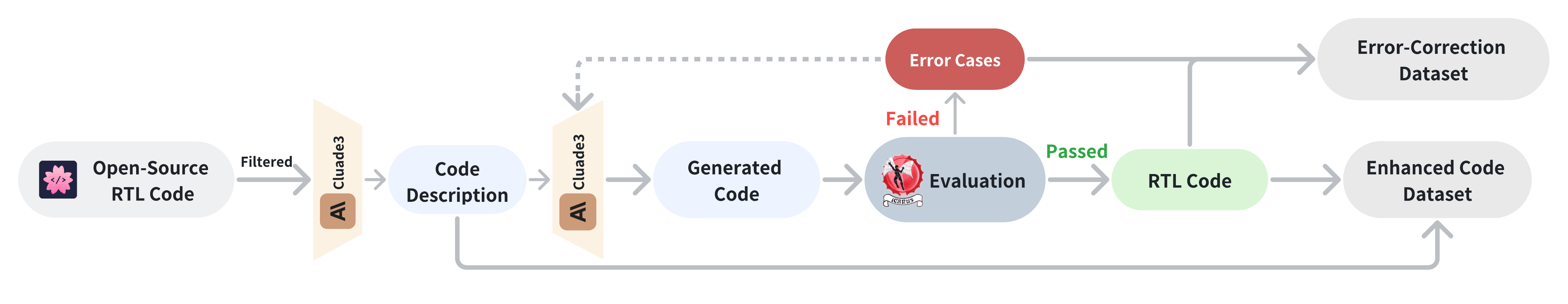}
    \caption{Code-to-Code Augmentation}.
    \label{fig:code-to-code}
\end{figure*}

\begin{figure}[h]
    \centering
    \includegraphics[width=\linewidth]{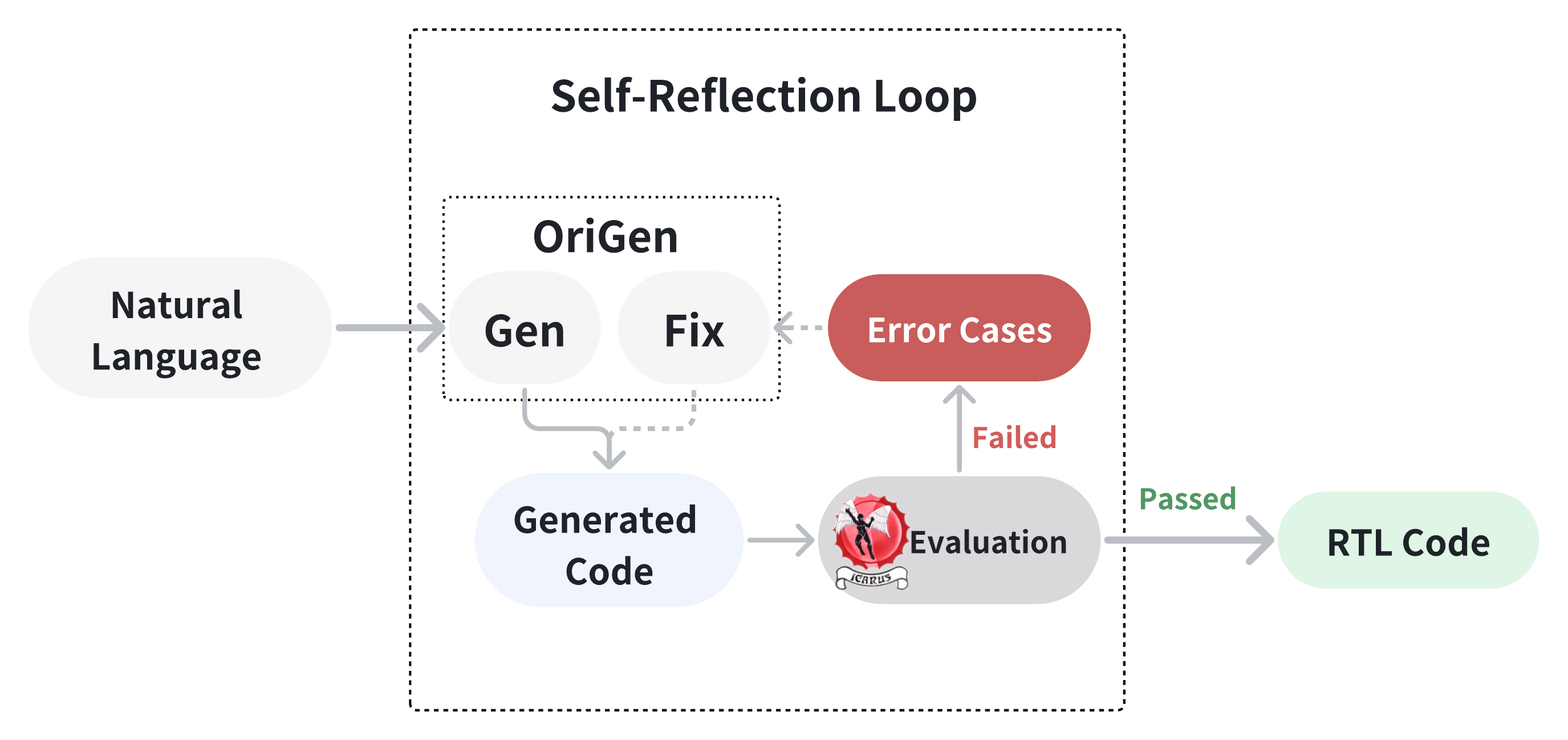}
    \caption{Generation and Self-Reflection}.
    \label{fig:generation}
\end{figure}

The overview of the code-to-code augmentation methodology and \origen's generation and self-reflection framework are illustrated in Figure \ref{fig:code-to-code} and Figure \ref{fig:generation}, respectively.

As shown in Figure \ref{fig:code-to-code}, our code-to-code augmentation process generates two datasets: an enhanced code dataset and an error-correction dataset.
The enhanced code dataset pairs Verilog code with natural language descriptions, enabling the LLM to learn the mapping between RTL code structures and high-level semantics. 
The error-correction dataset contains instructions erroneous, code samples and their corrections, enhancing the LLM's self-reflection and error correction capabilities.
This dataset is constructed by collecting cases where generated code fails evaluation, documenting the specific errors and compiler feedback, and providing corrected versions of the code that pass evaluation.

As shown in Figure \ref{fig:generation}, in the code generation and error correction stage, \origen comprises a base LLM and two trained LoRA  (Low-Rank Adaptation) models: Gen LoRA and Fix LoRA. 
Gen LoRA is fine-tuned on the enhanced code dataset. 
It specializes in interpreting natural language instructions and generating initial RTL code that aligns with the given specifications. 
Fix LoRA is fine-tuned on the error-correction dataset and built upon Gen LoRA. 
This component is responsible for analyzing compiler error messages, identifying code issues, and proposing and implementing fixes to erroneous code.
\origen receives natural language instructions as input, and Gen LoRA generates initial RTL code based on these instructions. 
The generated code then undergoes compilation evaluation. 
If the code fails to pass evaluation, it enters the self-reflection loop.
In this loop, Fix LoRA analyzes the error messages and the erroneous code, proposes and implements fixes to address the identified issues, and the corrected code is re-evaluated. 
This process iterates until the code passes verification or reaches a predefined maximum number of iterations.


\subsection{Code-to-Code Augmentation} \label{subsec:code-to-code}
The code-to-code augmentation process aims to transfer advanced RTL code generation and correction capabilities from commercial LLMs to our model. 
To achieve this, a carefully filtered collection of comprehensive open-source RTL code samples is utilized as a foundation. 

To extract a valuable dataset from open-source RTL code samples ~\cite{stackv2, verigen}, a rigorous filtration process is applied.
Initially, due to the constraints imposed by the model's context window and the challenges associated with incomplete descriptions for longer code snippets, samples exceeding 300 lines or 1536 tokens are excluded. 

Additionally, samples with an average of more than 30 tokens (approximately 90 characters) per line are considered non-standard and are consequently eliminated, ensuring that only concise and standard code samples are retained.
Subsequently, to ensure the meaningfulness and substantive content of the code snippets, a keyword-based filtration approach is employed. 
Each sample must contain both the \texttt{module} and \texttt{endmodule} keywords, alongside at least one occurrence of keywords related to procedural blocks, \texttt{always} (inclusive of variants like \texttt{always\_comb}, \texttt{always\_ff}, \texttt{always\_latch}, \etc) or \texttt{assign}. 
This criterion guarantees that the selected snippets are representative of functional and logical hardware designs.
Lastly, all comments within the code samples are removed. This step is crucial to prevent extraneous information from influencing the generation of accurate and relevant specifications.

Following the implementation of the aforementioned filtering procedures, the closed-source LLM Claude3-Haiku is utilized to generate detailed descriptions corresponding to the filtered code samples as shown in Figure \ref{fig:code-to-code}. The prompt we use is illustrated in \autoref{fig:description prompt}.
These descriptions are then used to regenerate RTL code, which replaces the original code in the dataset.
During the regeneration process, the generated code undergoes verification using the open-source compiler Icarus Verilog (Iverilog) ~\cite{iverilog}. 
If the code fails to compile, the compiler's error messages and the erroneous code are fed back into the LLM for regeneration to fix error.
\begin{figure}[htbp]
\centering\small
\begin{tikzpicture}
\node (text) at (0,11) [draw, thick, fill=none, rounded corners,
minimum width=8cm, minimum height=1cm] {
\begin{minipage}{8cm}
    \vspace{1mm}
    \textcolor{blue}{Description Prompt} \\
    Explain the high-level functionality of the Verilog module.\\
    \textbf{Task:}\\
    Please analyze it and provide a detailed description of its signals and functionality.
    Use as many high-level concepts that are directly applicable to describe the code, 
    but do not include extraneous details that aren't immediately applicable. 
    Speak concisely as if this was a specification for a circuit designer to implement. You should only reply with descriptive natural language and not use any code.\\
    \textbf{Code:}\\
    \{code\}\\
    \textbf{Response:}
    \vspace{1mm}
    \end{minipage}
};
\end{tikzpicture}
\caption{Prompt for Generating Code Descriptions}
\label{fig:description prompt}
\end{figure}

Simultaneously, code samples that fail compilation are utilized as foundational elements for the error-correction dataset, as detailed in Section \ref{subsec:error-correction}. 
This iterative process is repeated until the code successfully compiles or until the maximum number of iterations is reached, further enhancing the dataset's quality. 
This method facilitates the transfer of knowledge from the commercial LLM to our model, thereby improving its capability to generate and correct RTL code efficiently.

\begin{figure}[h]
    \centering
    \includegraphics[width=\linewidth]{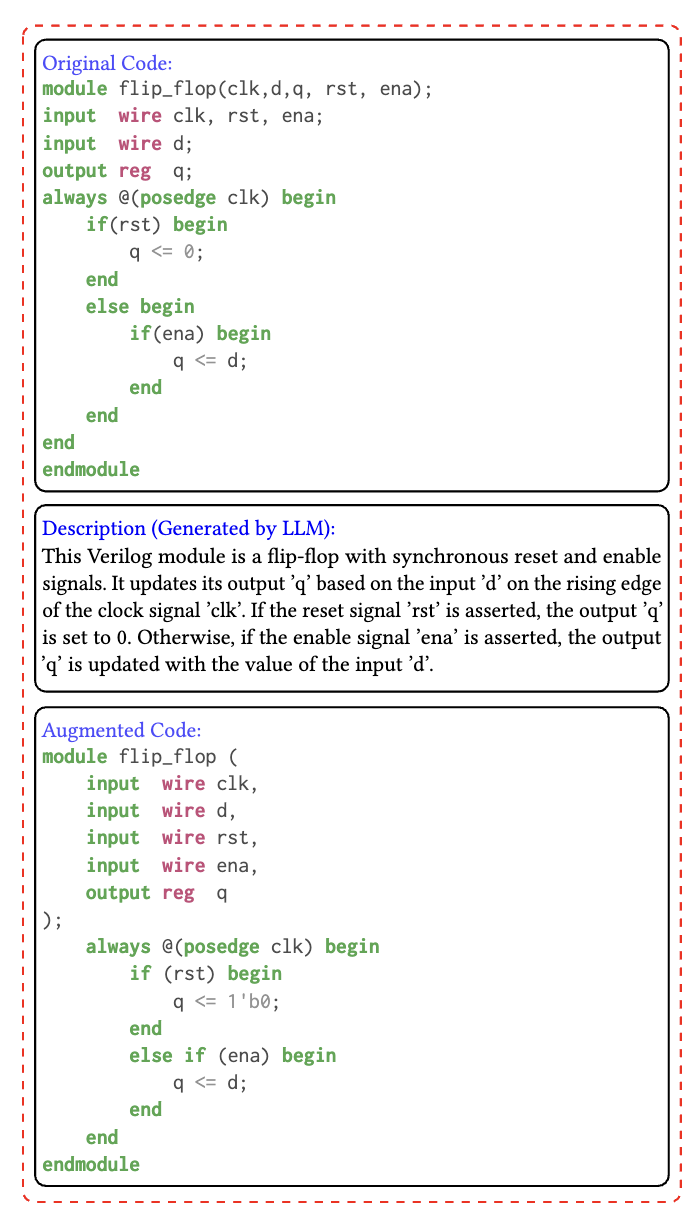}
    \caption{Example of Dataset Augmentation}
    \label{fig:example of dataset augmentation}
\end{figure}

A full example of dataset augmentation is shown in \autoref{fig:example of dataset augmentation}.
The process begins with the original Verilog code, which defines a flip-flop with synchronous reset and enable signals. 
The LLM generates a description of this code, explaining its functionality, specifically, how the output q is updated based on the input d on the rising edge of the clock clk. 
Using this description, the LLM produces an augmented version of the code. 
The enhanced code presents a more concise and organized structure while maintaining the same functionality, demonstrating the effectiveness of LLMs in improving RTL code readability and quality.

\subsection{Error-Correction Dataset} \label{subsec:error-correction}

Although \origen, after being trained on the enhanced code dataset, demonstrates performance comparable to that of advanced closed-source LLMs in RTL code generation, it exhibits weaker self-reflection capabilities compared to commercial LLMs like other open-source LLMs.

Given the observed performance gap, we decide to adopt the similar method to acquire stronger error understanding and self-reflection capabilities from the closed-source LLM. 
During the code-to-code augmentation process discussed in Section \ref{subsec:code-to-code}, the closed-source LLM generates a diverse set of RTL code samples that fail to compile, which are then corrected in the subsequent self-reflection process. 
We select the code samples generated in the code-to-code augmentation process that pass compilation after correction, along with the corresponding samples that failed to compile. 
The natural language instruction descriptions and compiler error messages associated with these samples serve as data sources for constructing the error-correction dataset.
This approach enables the model to be exposed to a broad spectrum of error types and their respective corrections, facilitating the development of robust error understanding and self-reflection capabilities.

Subsequently, these data samples are further filtered to ensure that the model learns to make modifications within the module's body rather than altering the module's declaration. 
Specifically, during the self-reflection and rectification process, the model may attempt to correct syntactic errors by modifying declarations to pass the compilation check though it is explicitly prohibited in the prompt. 
To prevent this, we remove samples with such modifications from the collected code.
This approach is crucial for preserving the overall structure and interface of the module, which is essential for maintaining compatibility with other modules in the design. 
Meanwhile, by restricting modifications to the module's body, we encourage the model to focus on identifying and correcting errors within the actual implementation logic. 
Through the above methods, we have successfully constructed a high-quality error-correction dataset.
This dataset comprises a large number of effectively repaired code samples generated by the Claude3-Haiku model, along with the corresponding erroneous code, compiler error messages, and natural language instructions.

To verify the effectiveness of the error-correction dataset, we introduce a benchmark named VerilogFixEval detailed in Section \ref{subsec:verilogfixeval}. 
\subsection{Code Generation and Fix}

\begin{figure}[h]
    \centering
    \includegraphics[width=\linewidth]{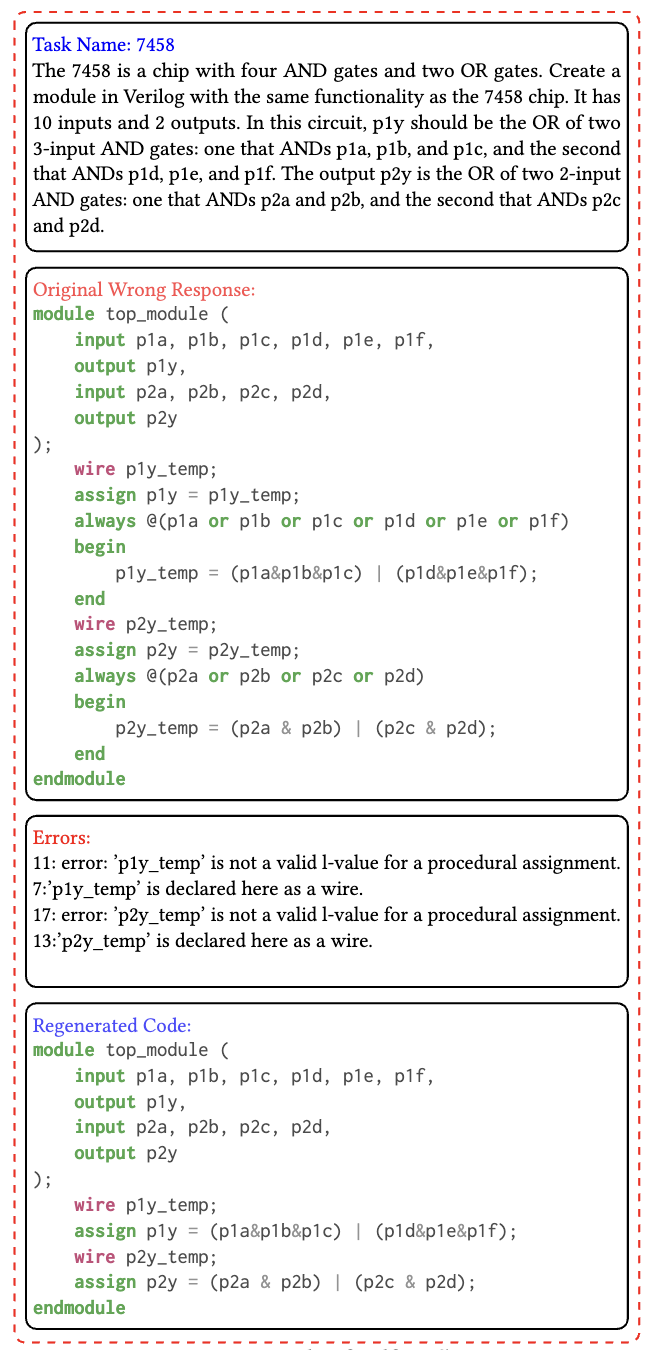}
    \caption{Example of Self-Reflection}
    \label{fig:debug_example}
\end{figure}

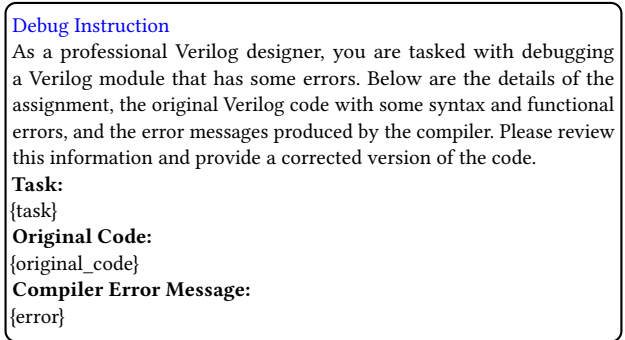
\begin{figure}
\centering\small
\begin{tikzpicture}
\node (text) at (0,11) [draw, thick, fill=none, rounded corners,
minimum width=8cm, minimum height=1cm] {
\begin{minipage}{8cm}
    \vspace{1mm}
    \textcolor{blue}{Debug Instruction} \\
    As a professional Verilog designer, you are tasked with debugging a Verilog module that has some errors. Below are the details of the assignment, the original Verilog code with some syntax and functional errors, and the error messages produced by the compiler. Please review this information and provide a corrected version of the code.\\
    \textbf{Task:}\\
    \{task\}\\
    \textbf{Original Code:}\\
    \{original\_code\} \\
    \textbf{Compiler Error Message:}\\
    \{error\}
    \vspace{1mm}
    \end{minipage}
};
\end{tikzpicture}
\caption{Debug Instruction Template}
\label{fig:debug instruction template}
\end{figure}

As shown in \autoref{fig:generation}, in the code generation and error correction stage, \origen comprises a base LLM and two trained LoRA models: Gen LoRA and Fix LoRA.
Following training on the enhanced code dataset, Gen LoRA has developed robust RTL code generation capabilities but exhibits limitations in self-reflection. 
To further enhance its capabilities of self-reflection, Gen LoRA is trained on the error-correction dataset, resulting in Fix LoRA. 

The reason for employing two LoRA models, Gen LoRA and Fix LoRA, instead of a single LoRA is based on experimental results, which indicate that Fix LoRA, trained on the error-correction dataset, exhibits inferior performance compared to Gen LoRA in the RTL code generation task. 
This performance degradation may be attributed to the fact that training on a dataset containing syntactic errors could potentially weaken the model's overall performance. 
Therefore, \origen adopts a two-LoRA approach, where Gen LoRA is responsible for generating the initial code, and Fix LoRA is tasked with rectifying syntactic errors, utilizing its specialized training on the error-correction dataset.


Figure \ref{fig:debug_example} illustrates an example of self-reflection in the correction of Verilog code for a module that implements the functionality of the 7458 chip, which includes four AND gates and two OR gates. 

The generated code initially contains a syntactic error: wires `ply\_temp` and `p2y\_temp` are incorrectly assigned values within always blocks. 
In Verilog, wires cannot be assigned values on the left-hand side of procedural assignments. 
This error is corrected by using the Fix LoRA model during the self-reflection process, following the instruction template shown in \autoref{fig:debug instruction template}. 
The regenerated code resolves these issues by employing continuous assignments for `p1y` and `p2y`, rather than assigning values within always blocks, ensuring that the logic accurately reflects the intended operations.

\subsection{VerilogFixEval Benchmark} \label{subsec:verilogfixeval}
To assess the ability of various models to reflect and improve from Verilog compiler error messages, we developed the VerilogFixEval benchmark. 
This benchmark consists of code samples that failed compilation verification, along with corresponding natural language instructions and compiler error messages. 
The faulty RTL code samples were selected from those generated by LLMs that performed similarly to GPT-3.5 on the VerilogEval benchmark. 
This selection method was adopted because \origen achieves a relatively high compilation pass rate. 
By including code samples from poorly performing LLMs, the benchmark ensures a diversity of errors while minimizing potential bias in test results that could favor \origen due to errors produced by \origen itself.

During the evaluation process, the model will receive natural language instructions, faulty RTL code, and compiler error messages to correct the RTL code.
The final evaluation metrics consist of two components: syntactic correctness and functional correctness.

\section{Evaluation}
\label{sec:evaluation}

\subsection{Experimental Setting}
\begin{table*}[h]
\centering
\caption{Comparison of functional correctness on VerilogEval~\cite{verilogeval} and RTLLM~\cite{rtllm}}
\label{tab:verillogeval}
\begin{tblr}{Q[m,c]|l|ccc|ccc|c}
\hline[1.2pt]
\SetRow{font=\bfseries}
\rii Source                     & \rii Name             & \ciii VerilogEval-human(\%)     &&& \ciii VerilogEval-machine(\%)   &&& RTLLM(\%) \\
\hline
\SetRow{font=\bfseries}
                                &                       & pass@1    & pass@5    & pass@10   & pass@1    & pass@5    & pass@10   & pass@5    \\
\hline
\rvi Commercial LLM             & GPT-3.5~\cite{gpt4}               & 35.6      & 48.8      & 52.6      & 49.4      & 72.7      & 77.6      & 44.8      \\
                                & GPT-4 2023-06-13~\cite{gpt4}      & 43.5      & 55.8      & 58.9      & 60.0      & 70.6      & 73.5      & \hii65.5      \\
                                & GPT-4 Turbo 2024-04-09~\cite{gpt4}& \hiii54.2  & \hi68.5   & \hi72.4   & 58.6      & 71.9      & 76.2      & \hii65.5      \\
                                & Claude3-Haiku~\cite{claude3}         & 47.5      & 57.7      & 60.9      & \hiii61.5      & 75.6      & 79.7      & 62.1      \\
                                & Claude3-Sonnet~\cite{claude3}        & 46.1      & 56.0      & 60.3      & 58.4      & 71.8      & 74.8      & 58.6      \\
                                & Claude3-Opus~\cite{claude3}          & \hi54.7   & \hii63.9  & \hii67.3      & 60.2      & 75.5      & 79.7      & \hi69.0    \\
\hline
\riii Open Source Models        & CodeLlama-7B-Instruct~\cite{codellama} & 18.2      & 22.7      & 24.3      & 43.1      & 47.1      & 47.7      & 34.5      \\
                                & CodeQwen1.5-7B-Chat~\cite{qwen}   & 22.4      & 41.1      & 46.2      & 45.1      & 70.2      & 77.6      & 37.9      \\
                                & DeepSeek-Coder-7B-Instruct-v1.5~\cite{deepseek} & 31.7 &42.8  & 46.8      & 55.7      & 73.9      & 77.6      & 37.9      \\
\hline
\rv Verilog-Specific Models     & ChipNeMo~\cite{chipnemo}              & 22.4      & -         & -         & 43.4      & -         & -         & -         \\
                                & VerilogEval~\cite{verilogeval}           & 28.8      & 45.9      & 52.3      & 46.2      & 67.3      & 73.7      & -         \\
                                & RTLCoder-DeepSeek~\cite{rtlcoder}     & 41.6      & 50.1      & 53.4      & 61.2      & \hiii76.5      & \hiii81.8      & 48.3      \\
                                & CodeGen-6B MEV-LLM~\cite{multi}    & 42.9      & 48.0      & 54.4      & 57.3      & 61.5      & 66.4      & -         \\
                                & BetterV-CodeQwen~\cite{betterv}      & 46.1      & 53.7      & 58.2      & \hii68.1      & \hii79.4      & \hii84.5      & -         \\
\hline
\cii \textbf{OriGen (ours)}     &                       & \hii54.4 & \hiii60.1     & \hiii64.2      & \hi74.1   & \hi82.4   & \hi85.7    & \hii65.5      \\
\hline[1.2pt]
\end{tblr}
\end{table*}
For our pre-trained model, we selected the DeepSeek-Coder-7B-Instruct model, as it exhibits the best performance in Verilog code generation among all 7B models, to the best of our knowledge. 
It is to ensure that our pre-trained model possesses strong capabilities in the domain of Verilog code generation, providing a solid foundation for further fine-tuning and evaluation.

To evaluate the performance of Verilog code generation, we selected two representative benchmarks: VerilogEval~\cite{verilogeval} and RTLLM~\cite{rtllm}. 
The former, VerilogEval, originated from approximately 150 Verilog tasks on the HDLBits website, which were manually converted to create VerilogEval-Human and generated by GPT-3.5 to produce VerilogEval-Machine. 
The latter benchmark, RTLLM, consists of 29 Verilog tasks with more diverse levels of difficulty, closely aligned with real-world design tasks. 
Both benchmarks employ the widely adopted pass@k evaluation metric to assess the correctness of the generated code's functionality. 
In this metric, if any one of the k samples passes the unit test, the problem is considered solved.
\begin{equation}
pass@k:=\underset{Problems}{\operatorname*{\mathbb{E}}}\left[1-\frac{\binom{n-c}k}{\binom nk}\right]
\end{equation}
where we generate $n \ge k$ samples for each instruction in which $c \le n$ samples pass testing. We choose $n = 10$ in experiments.

To assess the models' capability for self-reflection, we utilized the VerilogFixEval benchmark, as discussed in Section ~\ref{subsec:verilogfixeval}. This benchmark employs the pass@1 metric to evaluate the syntactic and functional accuracy of the rectified RTL code.

\subsection{Model Training}
We employ the LoRA (Low-Rank Adaptation) ~\cite{lora} method to train the model's capability in generating RTL code. 
This approach allows for enhancing the model's specific abilities in RTL code generation while minimizing the impact on its other capabilities. 
For all the training processes, we employ the float16 mixed precision method, although the model is trained in bfloat16 precision. 
We use the AdamW optimizer with parameters $\beta_1 = 0.9$ and $\beta_2 = 0.999$, along with cosine learning rate decay for scheduling. 
The warm-up ratio is set to 0.03, and the batch size is 8.

\subsection{Functional Correctness} \label{subsec:functional correctness}
Table \ref{tab:verillogeval} presents the results on the VerilogEval benchmark. 
To ensure fairness, we did not utilize the self-reflection feature for this comparison and generated code in a single attempt, like other models.
The models compared include closed-source commercial LLMs such as GPT-3.5/GPT-4, Claude3-Haiku/Sonnet/Opus, general open-source code models ~\cite{deepseek, qwen}, and models customized for RTL code generation ~\cite{chipnemo, verilogeval, rtlcoder, betterv, multi}.

In the VerilogEval benchmark ~\cite{verilogeval}, \origen achieves pass@1 scores of 54.4\% in the Human category and 74.1\% in the Machine category, outperforming remarkably all other alternatives designed for RTL code generation.
For instance, compared to the best-performing fully open-source RTLCoder~\cite{rtlcoder}, \origen surpasses it by 12.8\% on the Human benchmark.
Additionally, compared to commercial closed-source models, \origen demonstrates exceptional performance. 
It significantly outperforms GPT-4 (2023), which was previously considered a key benchmark. 
Furthermore, it exceeds the teacher model Claude3-Haiku, underscoring the effectiveness of filtering and self-reflection in code augmentation.

Among the state-of-the-art models, \origen outperforms Claude3-Haiku/Sonnet across all metrics and surpasses GPT-4 Turbo in the pass@1 metric while being only slightly inferior to Claude3-Opus by less than 1\% on the Human benchmark pass@1 metric.
Moreover, as shown in Table \ref{tab:verillogeval}, it achieves the best performance on the Machine benchmark, notably outperforming other models including Claude3-Opus and GPT-4 Turbo.
The performance gap between Human and Machine categories primarily stems from descriptions generated by GPT-3.5 for the Machine category. 
In contrast, \origen is trained on a vast amount of synthesized data, enabling it to excel with problems generated by LLMs.

\begin{figure}[t]
    \centering
    \includegraphics[width=\linewidth]{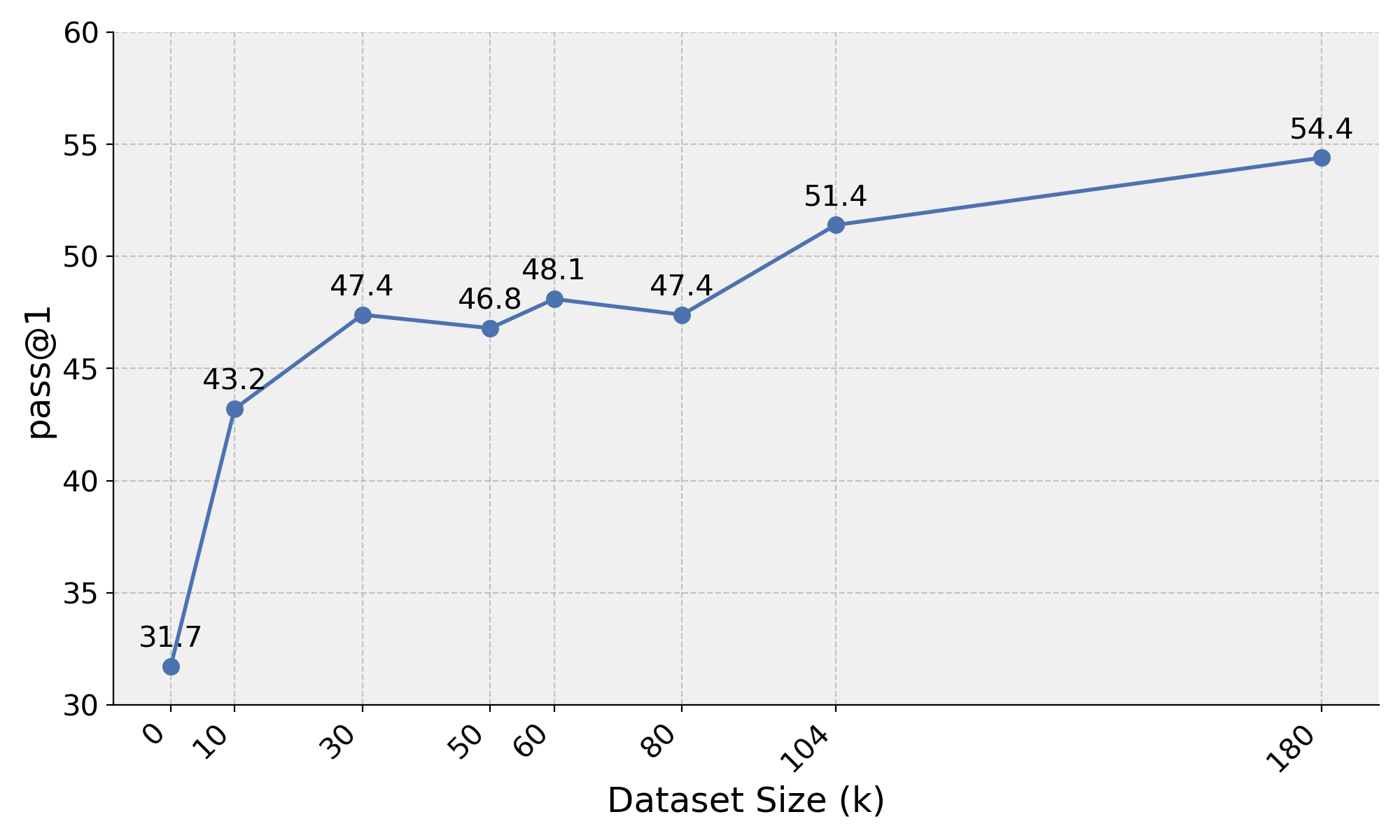}
    \caption{Pass@1 Variation with Dataset Size}
    \label{fig:dataset size}
\end{figure}
We also conduct experiments to investigate the relationship between dataset size and model performance, as depicted in Figure\ref{fig:dataset size}. 
The performance is evaluated using the pass@1 metric on VerilogEval Human. 
We examined dataset sizes ranging from 0 (representing the base model without fine-tuning) to approximately 180,000 samples (utilizing the entire fine-tuning dataset).

The results demonstrate a clear positive correlation between dataset size and model performance. 
Starting from a pass@1 score of 31.7 for the base model, we observe a sharp initial increase to 43.2 with just 10k samples. 
This is followed by a more gradual but consistent improvement as the dataset size increases, eventually reaching a peak performance of 54.4 with the full dataset of 180k samples.
Notably, the performance gains are not linear. 
The most substantial improvements occur in the early stages of data addition, with diminishing returns as the dataset size approaches its maximum. 
This pattern underscores the critical role of data quantity and diversity in model performance for further improvements beyond a certain scale.

To prevent our model from over-fitting on the VerilogEval benchmark, another 
benchmark RTLLM ~\cite{rtllm} is also used to evaluate.
Similar results are observed on RTLLM as shown in Table \ref{tab:verillogeval}, where \origen significantly outperforms other models, achieving performance comparable to GPT-4 Turbo and Claude3-Opus.

In summary, \origen significantly outperforms all non-commercial models across all metrics on both benchmarks and achieves comparable performance to the current state-of-the-art commercial closed-source models.

\subsection{Capability of Self-Reflection} \label{subsec:capability of self-reflection}
The evaluation metrics of VerilogFixEval consist of two components: syntactic correctness and functional correctness. 
Syntactic correctness assesses whether the generated rectified RTL code successfully compiles. 
Functional correctness further evaluates whether the generated rectified RTL code passes simulation tests.

As illustrated in Table \ref{tab:verilogfixeval}, experimental results on VerilogFixEval demonstrate that \origen achieves the best performance in syntactic error correction, surpassing GPT-4 Turbo by 19.9\%. 
This indicates that after training on the code-correction dataset, the model acquires powerful self-reflection and code rectification capabilities. 
For functional correctness, \origen significantly outperforms RTLCoder by 18.2\% and falls short of the best-performing model, Claude3-Opus, by only 6.3\%. 
This showcases \origen's ability to consider the functional correctness of the generated code while rectifying syntactic errors.

\begin{table}
\centering\small
\caption{Comparison of Models on VerilogFixEval}
\label{tab:verilogfixeval}
\begin{tblr}{Q[m,l]X[m,c]X[m,c]}
\hline[1.2pt]
\SetRow{font=\bfseries}
Model                                   & Syntactic correctness(\%) & Functional correctness(\%)\\
\hline
CodeQwen1.5-7B-Chat~\cite{qwen}                     & 27.6                      & 10.4                      \\
RTLCoder~\cite{rtlcoder}                                & 46.6                      & 15.3                      \\       
DeepSeek-Coder-7B~\cite{deepseek}                       & 50.7                      & 19.0                      \\
GPT-3.5~\cite{gpt4}                                 & 40.7                      & 13.1                      \\
GPT-4 Turbo 2024-04-09~\cite{gpt4}                  & 69.2                      & 37.1                      \\
Claude3-Haiku~\cite{claude3}                           & 48.9                      & 23.5                      \\
Claude3-Opus~\cite{claude3}                                  & 71.9                      & \h39.8                    \\
\hline
\textbf{\origen(Ours)}                  & \h89.1                    & 33.5                      \\
\hline[1.2pt]
\end{tblr}
\end{table}

\subsection{Ablation Studies}

We perform two ablation experiments to investigate the efficacy of the code-to-code augmentation method and to examine the model's self-reflection capability before and after training on the error-correction dataset. For the ablation study of the code-to-code data augmentation method, we compared the performance of models trained on RTL code dataset before and after code-to-code augmentation method.
\begin{table}
\centering
\caption{Ablation Study of Code-to-Code Augmentation}
\label{tab:ablation augmentation}
\SetTblrInner{rowsep=1pt,colsep=2pt}
\begin{tblr}{Q[m,l]Q[m,c]Q[m,c]Q[m,c]}
\hline[1.2pt]
\SetRow{font=\bfseries}
Benchmark           & Metric    & Baseline(\%)  & Augment(\%)\\
\hline
VerilogEval-Human~\cite{verilogeval}         &   Pass@1  & 41.6          & 54.4       \\
VerilogEval-Machine~\cite{verilogeval} &   Pass@1  & 62.5          & 74.1       \\       
RTLLM~\cite{rtllm}               &   Pass@5  & 41.4          & 65.5       \\
\hline[1.2pt]
\end{tblr}
\end{table}
\begin{table}
\centering
\caption{Ablation Study of Error-Correction Dataset on VerilogFixEval}
\label{tab:ablation debug}
\SetTblrInner{rowsep=1pt,colsep=2pt}
\begin{tblr}{Q[m,l]X[m,c]X[m,c]}
\hline[1.2pt]
\SetRow{font=\bfseries}
Model & Syntactic Correctness(\%) & Functional Correctness(\%)\\
\hline
baseline & 53.8 & 23.4 \\
baseline + error message & 63.5 & 25.6 \\
finetune & 82.7 & 31.9 \\
finetune + error message & 89.1 & 33.5 \\
\hline[1.2pt]
\end{tblr}
\end{table}
Table \ref{tab:ablation augmentation} demonstrates that across multiple evaluation metrics, the model trained on the dataset enhanced by code-to-code augmentation significantly outperforms the model trained on the unaugmented dataset. 
This indicates that our model's outstanding performance is primarily attributed to the code-to-code augmentation methodology, as training on the unaugmented dataset does not yield substantial improvements.

For the ablation study of the error-correction dataset, the results are presented in Table \ref{tab:ablation debug}. 
We evaluate four scenarios, considering the model's performance before and after training on the error-correction dataset, as well as with and without the use of error messages from the compiler.
The former is denoted as "baseline" and "finetune," indicating whether the model is trained on the error-correction dataset, while the latter is represented by "error message."

The results demonstrate that after training on the error-correction dataset, the model significantly outperforms its pre-training performance in both syntactic correctness and functional correctness.
Additionally, the trained model exhibits a reduced reliance on error messages from the compiler.

\section{conclusion}
This paper introduces \origen, an open-source framework for RTL code generation. 
It proposes a novel code-to-code augmentation methodology to generate high-quality, large-scale RTL code datasets, which enhances the model's training data. 
The framework also introduces a self-reflection mechanism that allows \origen to autonomously fix syntactic errors by leveraging compiler feedback, thereby improving its code generation accuracy.
Furthermore, we construct a dataset to improve the model's capability of self-reflection based on compiler error messages and erroneous code and develop a benchmark to evaluate this capability. 
Experimental results demonstrate that \origen remarkably outperforms other open-source alternatives in RTL code generation, surpassing the previous best-performing LLM by 12.8\% on the VerilogEval-Human benchmark and is comparable with GPT-4 Turbo.
Moreover, \origen exhibits superior capabilities in self-reflection and error rectification, surpassing GPT-4 by 19.9\% in syntactic correctness on the VerilogFixEval benchmark.

\bibliographystyle{ACM-Reference-Format}
\bibliography{references}

\end{document}